\documentstyle[11pt,epsfig]{article}
\begin{document}
\begin{sloppypar}



\begin{center}

{\large \bf CORRELATION ANALYSIS OF THE INSTANTON-INDUCED PARTICLE 
PRODUCTION IN QCD \\ }

\vspace{15pt}
{\large V.Kuvshinov\footnote{kuvshino@dragon.bas-net.by}
and
\underline{R.Shulyakovsky}\footnote{shul@dragon.bas-net.by}}\\

{\it Institute of Physics \\
National Academy of Sciences of Belarus \\
Scarina av.,68, Minsk 220072 \\
BELARUS }

\vspace{15pt}
\end{center}
\noindent

Factorial, cumulant and
$H_q$-moments in dependence on their rank $q$ for the
{\it instanton-induced} deep inelastic scattering (DIS)
in the frameworks of QCD
are calculated and analysed.
The obtained correlation moments behaviour has specific form,
which can be considered as a new criterion of the
QCD-instantons identification on experiment at HERA.

\section{Introduction}
As it is known, such gauge theories as SM of electroweak interactions and
QCD have degenerated vacuum structure on the {\it classical}
level~\cite{JR}: potential energy is periodic with respect to the
Chern-Simons number

\begin{equation} \label{cs}
N_{cs}=\frac{g^2}{16\pi^2}\int d^3x\varepsilon_{ijk}
\Bigl( A^a_i\partial_jA^a_k+\frac{g}{3}
\varepsilon^{abc}A^a_iA^b_jA^c_k\Bigr).
\end{equation}

Minimal energy (classical vacua) corresponds
to integer $N_{cs}$. Neighbouring vacua are separated by a potential
barrier of height $E_{sp}$ (Fig.1).

\vspace{2.7cm}
\begin{center}
\begin{minipage}{5in}
\epsfxsize=3.5in \epsfysize=2in
\epsfbox{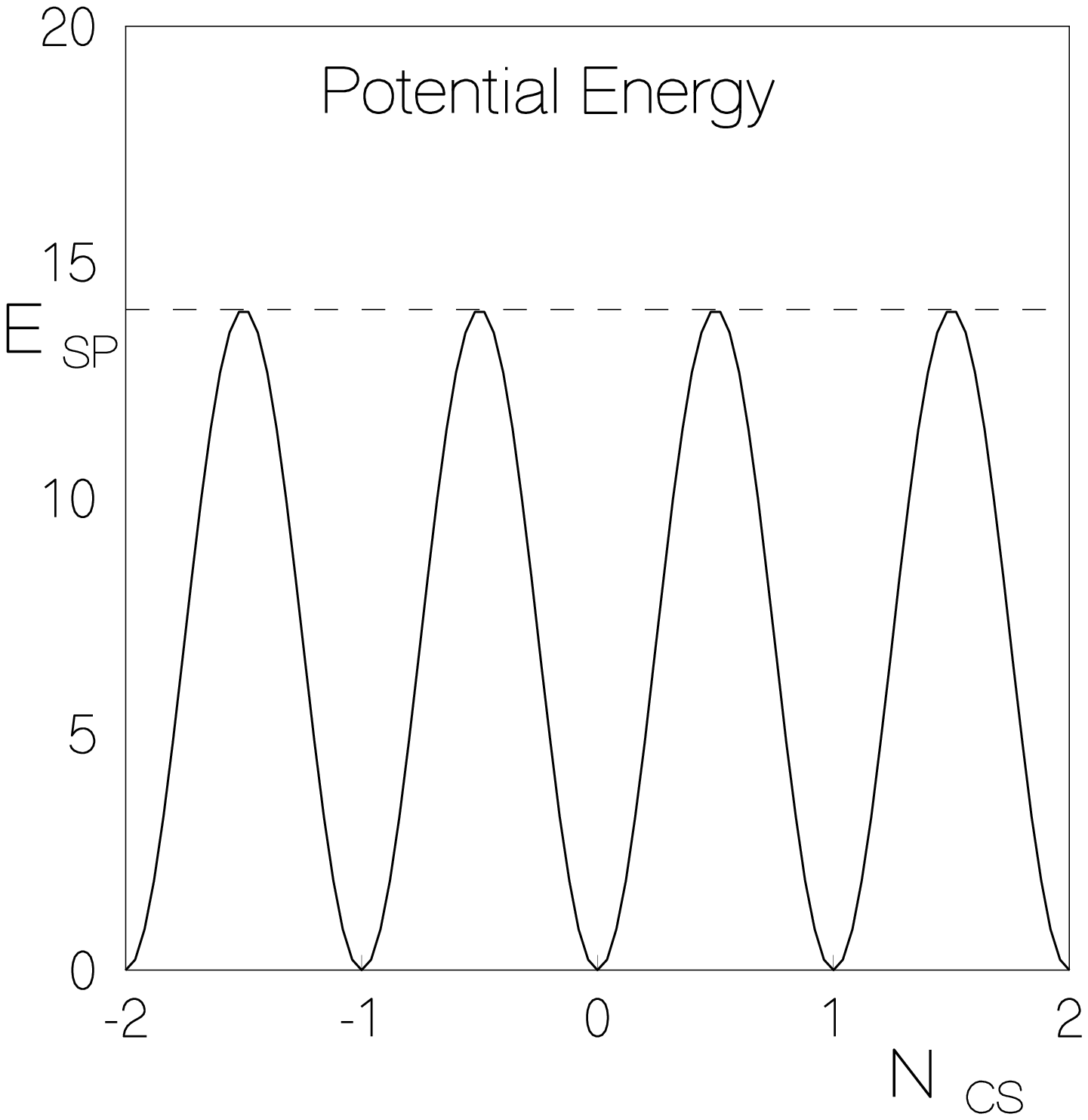}
\end{minipage}
\end{center}

\vspace*{-1.2cm}
\begin{minipage}{15cm}
Fig.1. Schematic dependence of potential energy of gauge fields on
Chern-Simons number $N_{cs}$. Gauge condition $A_0=0$ is used.
$E_{sp}$ is so-called sphaleron mass.
\end{minipage}
\vspace*{0.2cm}

Usual perturbative theory (Feynman rules) describes
phenomena with $N_{cs}=0$ only. {\it Quantum} tunnelling transitions
between neighbouring vacua can be described by means of {\it instantons},
which are classical solutions of the Euclidean field equations
with finite action~\cite{BPST}.
Taking into account such tunnelling transitions leads to the baryon
number violation in SM~\cite{Hooft}, which is connected with
the problem of matter and antimatter asymmetry in the Universe~\cite{Zuccero}.
In QCD instantons lead to the chirality violation, allow to solve
U(1)-problem~\cite{Hooft}, give contribution to the confinement~\cite{Pol}.
Therefore, the experimental discovery of instantons would be of
fundamental significance for particle physics.

It was suggested probability of the instanton transitions can
increase in high energy collisions~\cite{R90}.
There is a possibility of the
instanton-induced events identification in the
electron-proton DIS at HERA (DESY)~\cite{Bal}.
Instanton induced DIS final states can be distinguished from ordinary
(perturbative) ones through some features:

\noindent
- {\it high multiplicity} (the average number of
partons $\sim 10$~\cite{Bal});

\noindent
- {\it isotropic} distribution of partons in the instanton rest
system and presence, practically, of
{\it all} light quarks (u, d, s)
in each events~\cite{Hooft};

\noindent
- specific behaviour of gluon structure functions~\cite{RSh}
and gluon correlation characteristics~\cite{Acta}.

In our report additional "footprints" of QCD-instantons
(factorial, cumulant and $H_q$-moments)
are studied.

\section{Multiplicity distribution of the instanton-induced final states}

In DIS instantons can appear in the
quark-gluon subprocesses (Fig.2).
The following usual designations are used:

\begin{equation}
Q^2=-q^2,\qquad x=\frac{Q^2}{2Pq}, \qquad
Q'^2=-q'^2,\qquad x'=\frac{Q'^2}{2pq'},
\end{equation}

\noindent
where transferred momentum square $Q^2$ and Bjorken variable $x$
describe total DIS process; $Q'^2$ and $x'$ characterise the {\it
insanton} subprocess.

\vspace*{0.5cm}
\hspace*{3cm}
\begin{minipage}{3.5in}
\epsfxsize=2.5in \epsfysize=2.5in
\epsfbox{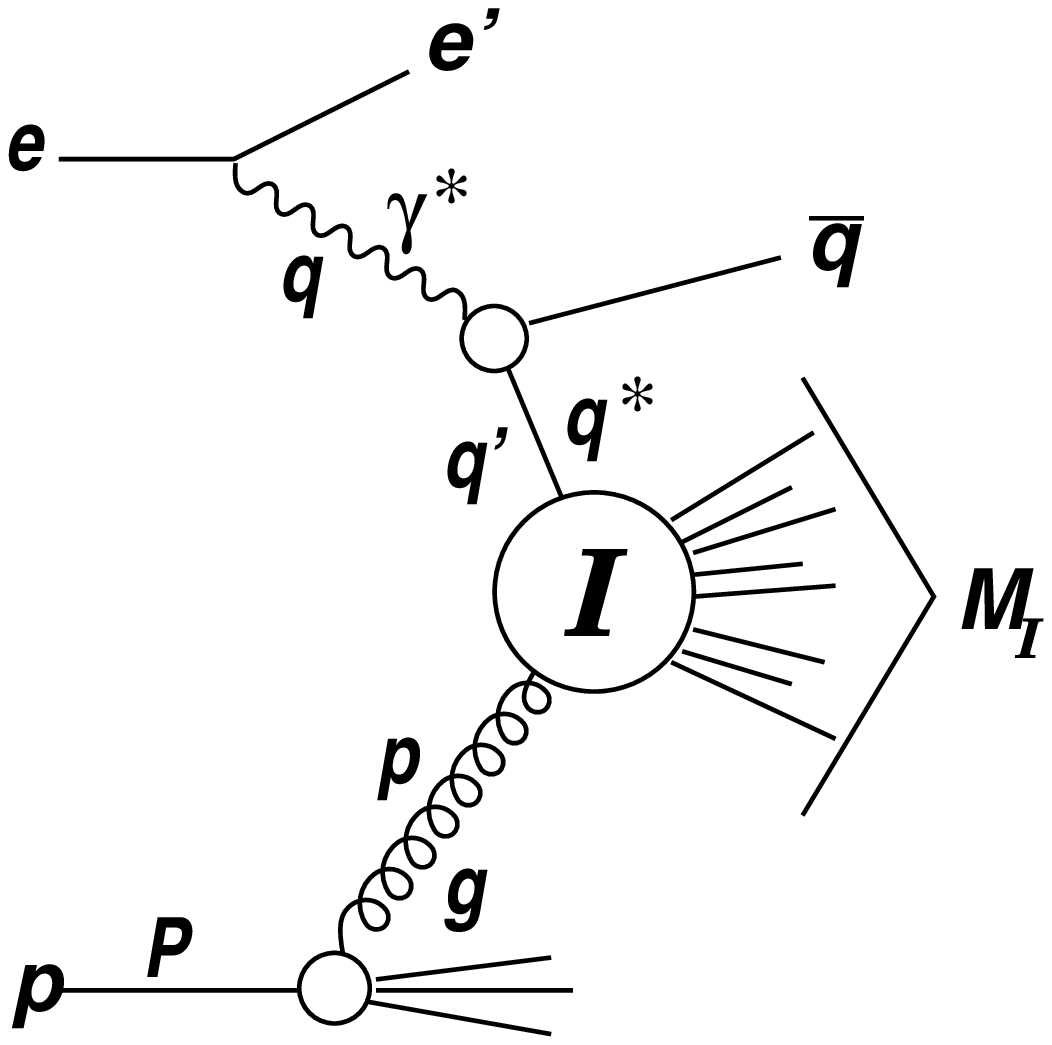}
\end{minipage}
\vspace*{-12.6cm}

\vspace*{13cm}
\begin{minipage}{16.5cm}
Fig.2. Instanton induced DIS (figure was taken from~\cite{Kuhlen}).
\end{minipage}
\vspace*{0.1cm}

As it was mentioned above high parton multiplicity is
one of the main characteristics of the
instanton-induced events.
The distribution on numbers of gluons in
the instanton-induced events is given by the expression:

\begin{equation}\label{glu}
P_n^{(g)}=\frac{1}{\sigma_{tot}}\frac{1}{n!}
\int d^4k_1...d^4k_n\Bigl|
T(k_1,...,k_n)\Bigr|^{2},
\end {equation}

\noindent
where $\sigma_{tot}$ -- total cross-section, $T(k_1,...,k_n)$
is the amplitude of the production
of gluons with the energy-momentum 4-vectors $k_1,...,k_n$.
It is calculated by means of LSZ-technique applying to
the Euclidean n-points Green function, which is given
by the following Feynman path integral (in the
quasiclassical approximation):

\begin{equation}\label{Fey}
\int DAe^{-S^e[A]}A^{I\ a_1}_{\ \mu_1}(x_1)...A^{I\ a_n}_{\ \mu_n}(x_n).
\end{equation}

\noindent
where $S^e[A]$ is QCD Euclidean action, $A^{I\ a}_{\ \mu}(x)$ --
instanton configuration~\cite{BPST}.
In quasiclassical approximation
Gauss integral (\ref{Fey}) is known calculable expression.
The integration is carried out on the gluon fields,
which connect neighbour classical vacua.
Factorisation in (\ref{Fey}) leads to the Poisson distribution
on the final gluon number~\cite{RSh, Acta}:

\begin{equation}\label{glu2}
P_n^{(g)}=e^{-<n_g>}\frac{<n_g>^n}{n!},\quad
<n_g>=\frac{16\pi^2}{g^2}\Biggr(\frac{1-x'}{x'}\Biggl)^2,\quad 0.5<x'<1.
\end{equation}

The quarks production in the instanton processes
is described by the
well-known fixed multiplicity distribution
(if we take into account zero modes only~\cite{Hooft}):

\begin{equation}\label{quark}
P_n^{(q)}=\delta_{2n_f,n},
\end{equation}

\noindent
where $n_f$ is a number of massless quark flavours.
We suggest that masses of u, d, s are equal to zero.

Thus, if we take into account both gluons and quarks,
then the following distribution is obtained:

\begin{equation}\label{both}
P_n=e^{-<n_g>}\frac{<n_g>^{n-2n_f}}{(n-2n_f)!}\Theta(n-2n_f).
\end{equation}

\section{ Calculation of the correlation moments for the instanton
DIS processes}

Study of the correlation moments is more useful sometimes
than study of the multiplicity distribution~\cite{Dremin}.
Let us remind the well-known definition of the normalised factorial
moments:

\begin{equation}
F_q=\frac{1}{<n>^q}\left. \frac{d^qQ(z)}{dz^q}\right|_{z=1},\qquad
Q(z)=\sum_{n=1}^{\infty }P_nz^n,\qquad
0\leq z\leq 1.
\end{equation}

\noindent
where $<n>$ is the average multiplicity, $Q(z)$ -- generating function.

In the case of the instanton-induced multiparticle production
processes $Q(z)$ and $<n>$ have the following forms:

\begin{equation}
Q(z)=\sum_{n=2n_f}^{\infty}e^{-<n_g>}\frac{<n_g>^{n-2n_f}z^n}{(n-2n_f)!}=
z^{2n_f}e^{<n_g>[z-1]},\quad
<n>=<n_g>+2n_f.
\end{equation}

\noindent
The corresponding normalised factorial moments dependence on $q$ is shown
on the Fig.3. It is well-known, that normalised factorial moments for
ordinary perturbative processes of the particle production increase with
increasing $q$.
Therefore the behaviour of the moments for the instanton-induced
processes can be used as a new instanton identification criterion.

Also we can consider the normalised cumulant moments:

\begin{equation}\label{k}
K_q=\frac{1}{<n>^q}\left. \frac{d^qlnQ(z)}{dz^q}\right|_{z=1}.
\end{equation}

\noindent
It is not difficult to calculate $K_q$ for the instanton distribution
(\ref{k}):

\begin{equation}\label{ki}
K_q=\frac{2n_f(-1)^{q-1}q!+<n_g>\delta_{q1}}{(<n_g>+2n_f)^q}.
\end{equation}

It is more interesting to consider the instanton contribution to the
ratio of the cumulant and factorial moments:

\begin{equation}\label{h}
H_q=\frac{K_q}{F_q}.
\end{equation}

\noindent
These moments have the following properties for the perturbative QCD:
decreasing oscillations, presence of the negative correlations,
there is the first minimum at $q=5$~\cite{Dremin}.

Unlike this, for the instanton distribution (\ref{both})
$H_q$-moments have the first minimum at
$q=2$ (Fig.4), oscillations, which magnitude increases at large $q$ numbers
(Fig.5).

\section{Conclusion}

The obtained dependences of the factorial, cumulant and $H_q$-moments
on their rank have specific forms.
Therefore, the behaviour of the correlation moments
can be used as a new signal of the QCD nonperturbative vacuum phenomenon
in addition to the well-known "footprints".
Of course, we need to
take into account hadronization stage. Local parton-hadron
duality~\cite{Azim} allows to apply the
obtained results for the experimental QCD-instantons search.

We propose the following procedure for the experimental QCD
instantons search at HERA:

\noindent
1) selection of the events with high multiplicity.
For our approximations $n_{hadrons}\ge 12$
(local parton-hadron duality gives
number of the final hadrons
$n_{hadrons}=2n_{partons}\ge 2n_{quarks}=4n_f=12$);

\noindent
2) analysis of the correlation moments for the {\it selected} events
and comparison with theoretical predictions.

\end{sloppypar}

\twocolumn
\vspace*{7cm}
\begin{minipage}{3in}
\epsfxsize=3in \epsfysize=3in
\epsfbox{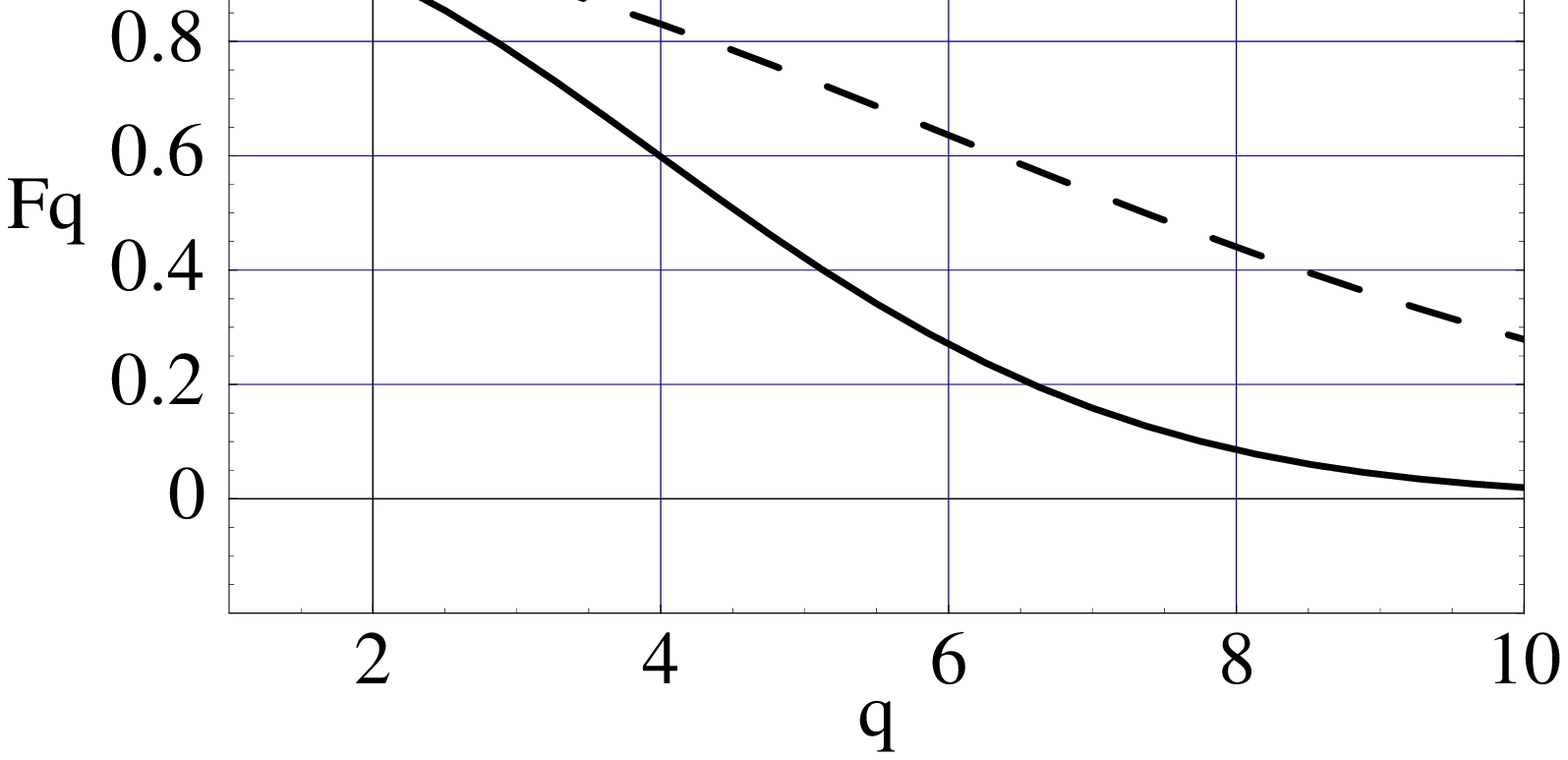}
\end{minipage}

\vspace*{-9cm}
\begin{minipage}{7.5cm}
Fig.3. The dependence of the normalised factorial moments on their rank
$q$ for the different average gluon numbers:
$<n_g>=2$ (solid line), $<n_g>=8$ (dash line); $n_f=3$.
\end{minipage}
\vspace*{0.2cm}

\vspace*{2.1cm}
\begin{minipage}{3in}
\epsfxsize=3in \epsfysize=3in
\epsfbox{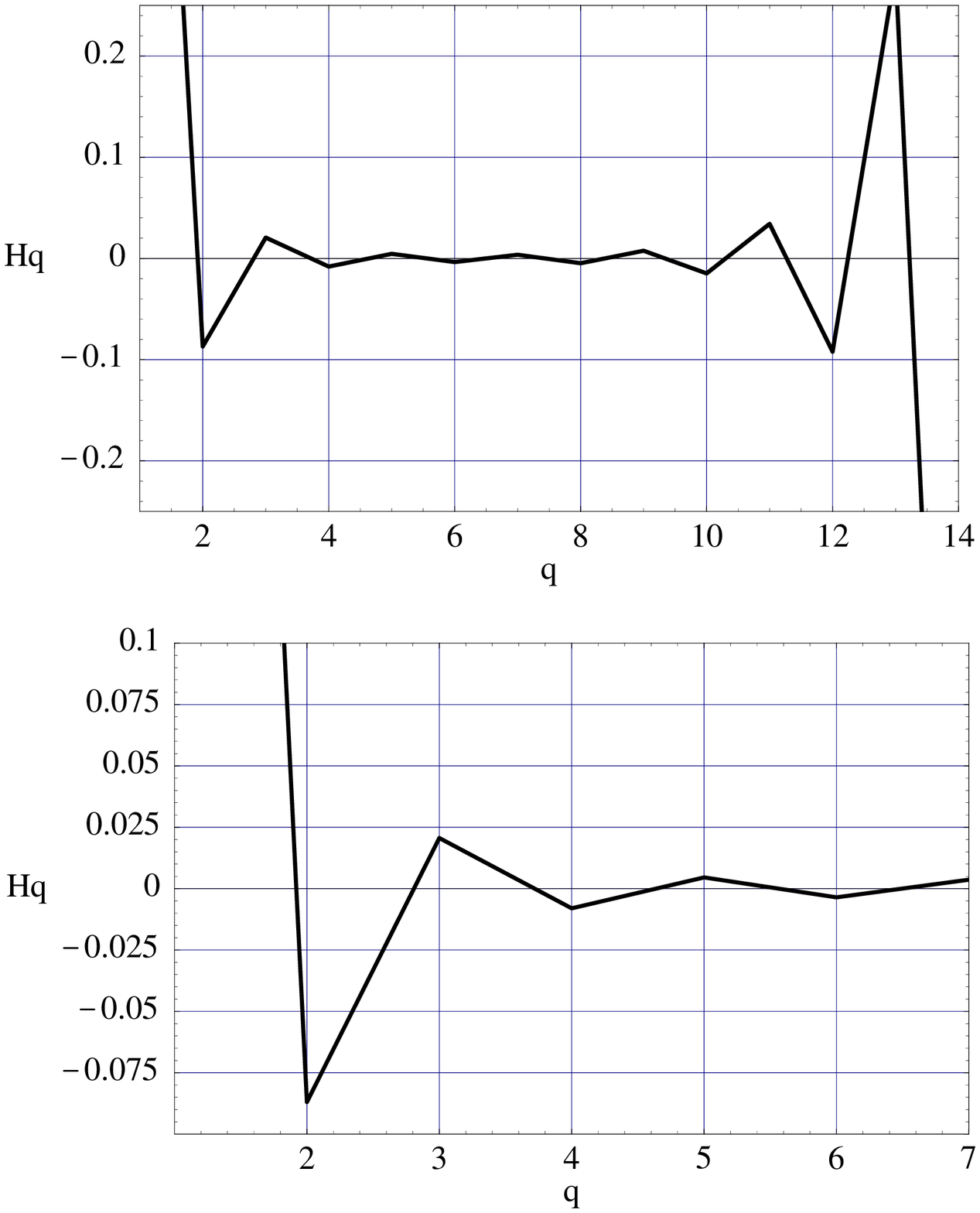}
\end{minipage}

\vspace*{0.1cm}
\begin{minipage}{7.5cm}
Fig.4. $H_q$ as the function of $q$,
$<n_g>=2$, $n_f=3$:
upper curve -- full plot,
lower -- first part of the plot.
\end{minipage}
\vspace*{0.7cm}

\newpage
\vspace*{2.3cm}
\begin{minipage}{5in}
\epsfxsize=3in \epsfysize=3in
\epsfbox{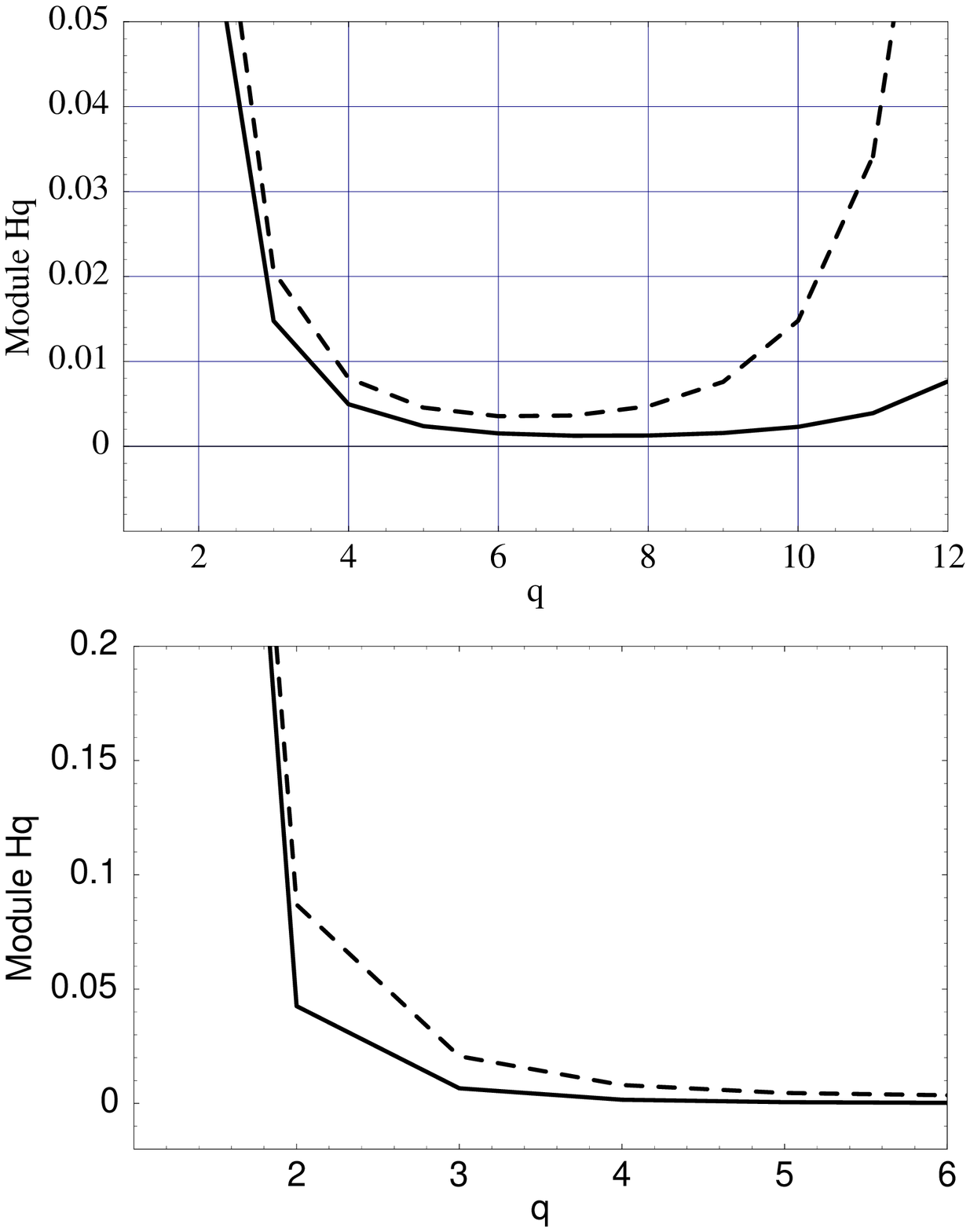}
\end{minipage}

\vspace*{0.3cm}
\begin{minipage}{7.5cm}
Fig.5. Absolute values of $H_q$ as the function of $q$.
Upper curve corresponds to the full plot
($<n_g>$=3 for solid line, $<n_g>=2$ for dash line),
lower -- first part of the plot ($<n_g>$=6 for solid line,
$<n_g>=2$ for dash line).
\end{minipage}
\vspace*{3.2cm}


\begin{thebibliography}{ab}

\bibitem{JR}
R.Jackiw and C.Rebbi, {\it Phys.\/Rev.\/Lett.}\/ {\bf 37} (1976) 172.
\bibitem{BPST}
A.Belavin, A.Polyakov, A.Schwarz and Yu.Tyupkin, {\it Phys.\/Lett.}\/
{\bf B59} (1975) 85.
\bibitem{Hooft}
G.'t Hooft, {\it Phys.\/Rev.\/Lett.}\/ {\bf 37} (1976) 8, \\
G.'t Hooft, {\it Phys.\/Rev.}\/ {\bf D14} (1976) 3432.
\bibitem{Zuccero}
A.Sakharov, {\it JETP Lett.}\/ {\bf 5} (1967) 1.
\bibitem{Pol}
A.Polyakov, {\it Phys.\/Lett.}\/ {\bf B59} (1975) 82.
\bibitem{R90}
A.Ringwald, {\it Nucl.\/Phys.}\/ {\bf B330} (1990) 1.
\bibitem{Bal}
I.Balitsky and V.Braun, {\it Phys.\/Lett.}\/ {\bf B314} (1993) 237.
\bibitem{RSh}
S.Moch, A.Ringwald, F.Schrempp,
{\it Nucl.\/Phys.}\/ {\bf B507} (1997) 134.
\bibitem{Acta}
V.Kuvshinov and R.Shulyakovsky, {\it Acta Phys.\/Pol}\/ {\bf B28} (1997) 1629,
hep-ph/9902403; \\
V.Kuvshinov and R.Shulyakovsky, {\it Acta Phys.\/Pol}\/
{\bf B30} (1999) 69, hep-ph/9902379.
\bibitem{Kuhlen}
T.Carli, M.Kuhlen, {\it DESY preprint}\/ {\bf 97-151} (August,1997),
hep-exp/9708008.
\bibitem{Dremin}
I.Dremin, {\it Phys. Rev. Lett.\/}\/ {\bf 313} (1993) 209; \\
I.Dremin, {\it Proceed. of the 7th Work\-shop on
Multi\-particle Production "Correlations and Fluctuations"
(June 30 - July 6, 1996, Nijmegen, The Netherlands)}\/, (1997) 313.
\bibitem{Azim}
Ya.Azimov, Yu.Dokshitzer, V.Khoze, S.Troyan,
{\it Z.\/Phys.}\/ {\bf C27} (1985) 65.

\end{thebibliography}
\end{document}